\documentclass[prl,twocolumn,superscriptaddress,floatfix]{revtex4-1}

\usepackage{graphicx}
\usepackage{multirow}

\begin{document}

\title{Engineering polarization rotation in a ferroelectric superlattice}
\author{J. Sinsheimer}
\author{S.J. Callori}
\author{B. Bein}
\author{Y. Benkara}
\author{J. Daley}
\author{J. Coraor}
\affiliation{Department of Physics and Astronomy, Stony Brook University, Stony Brook, NY 11794-3800 USA}
\author{D. Su}
\affiliation{Center for Functional Nanomaterials, Brookhaven National Laboratory, Upton, NY 11973-5000 USA}
\author{P.W. Stephens}
\affiliation{Department of Physics and Astronomy, Stony Brook University, Stony Brook, NY 11794-3800 USA}
\affiliation{Photon Sciences Directorate, Brookhaven National Laboratory, Upton, NY 11973-5000 USA}
\author{M. Dawber}
\email{matthew.dawber@stonybrook.edu}
\affiliation{Department of Physics and Astronomy, Stony Brook University, Stony Brook, NY 11794-3800 USA}

\begin{abstract}
A key property that drives research in ferroelectric perovskite oxides is their strong piezoelectric response in which an electric field is induced by an applied strain, and vice-versa for the converse piezoelectric effect. We have achieved an experimental enhancement of the piezoelectric response and dielectric tunability in artificially layered epitaxial PbTiO$_{3}$/CaTiO$_{3}$ superlattices through an engineered rotation of the polarization direction.  As the relative layer thicknesses within the superlattice were changed from sample to sample we found evidence for polarization rotation in multiple x-ray diffraction measurements. Associated changes in functional properties were seen in electrical measurements and piezoforce microscopy. The results demonstrate a new approach to inducing polarization rotation under  ambient conditions in an artificially layered thin film.

\end{abstract}

\maketitle

In ferroelectric perovskite oxides  \cite{DRS05} enhanced piezoelectric responses can occur when the direction of the polarization rotates due to some small change in a parameter, such as an applied electric field or stress. For example, the solid solution lead zirconate titanate, PbZr$_{1-x}$Ti$_{x}$O$_{3}$ (PZT) has greatly enhanced piezoresponse in a narrow range of composition of about $x=0.48$ \cite{Jaffe54}. In epitaxial thin films of ferroelectric materials, strain from the substrate on which they are grown allows for strain engineering, which in some cases leads to stunning results \cite{Haeni04,Choi04,Lee10}. Here we built upon the strain engineering approach and went a step further by engineering polarization rotation in artificially layered epitaxial superlattices. In addition to the strain control applied by the substrate, the properties of the structures can be engineered by making use of electrostatic interactions  that couple the constituent layers \cite{DawberPRL05,DawberAM07} or interactions between different modes at the interfaces between those layers \cite{Bousquet08}. Based on the expectation that in short period superlattices the polarization should be approximately continuous \cite{Neaton03,DawberPRL05}, and that piezoelectric strains should be equally shared \cite{Jo10} from one component layer to the next, it is interesting to consider whether a rotation of the polarization through one or more intermediate phases is achievable by varying the relative thicknesses of the constituent layers of the superlattice.

A key factor in the enhanced piezoresponse of PZT is the existence of a monoclinic phase which, as a function of composition, lies between the tetragonal phase and rhombohedral phase of this solid solution  \cite{Noheda99,Guo00}. Monoclinic phases have also been found to be crucial to enhanced piezoresponse in other solid solutions, such as Pb(Zn$_{1/3}$Nb$_{2/3}$)O$_{3}$-PbTiO$_{3}$ (PZN-PT) \cite{Noheda01} and Pb(Mg$_{1/3}$Nb$_{2/3})$O$_{3}$-PbTiO$_{3}$  (PMN-PT) \cite{Noheda02}. First principles calculations \cite{Fu00}, extensions to the phenomenological theory of ferroelectrics \cite{Vanderbilt01} and model Hamiltonian approaches \cite{Bellaiche01}, have revealed that the monoclinic phases allow pathways for the polarization to rotate and that this is the origin of the enhanced piezoresponse in these materials. Three different kinds of monoclinic phases are possible,  $M_{C}$ in which the polarization is directed along $[u0v]$, and $M_{A}$ and $M_{B}$ in which polarization is along $[uuv]$ with $u<v$ for  $M_{A}$  and $u>v$ for $M_{B}$.  The phase observed in PZT is the $M_{A}$ phase, while that in PZN-PT and PMN-PT is the $M_{C}$ phase \cite{Noheda01,Noheda02}. A more recent study has shown that as increasingly large pressure is applied to  pure PbTiO$_{3}$, the crystal undergoes consecutive transitions from $T-M_{C}-M_{A}-R$ \cite{Ahart08} (where $R$ is rhombohedral phase with polarization along [111]).

Due to the compressive strain of $-1.64\%$ that SrTiO$_{3}$ exerts on PbTiO$_{3}$,  PbTiO$_{3}$ epitaxially constrained in-plane to the lattice parameter of a SrTiO$_{3}$ substrate is tetragonal and has polarization directed along the $[001]$ direction.  By contrast CaTiO$_{3}$ on SrTiO$_{3}$ is under $2\%$ tensile strain. Based on first-principles calculations it is expected that the ground state of CaTiO$_{3}$ under sufficient tensile strain ($>1.5\%$) is orthorhombic ($O$) and ferroelectric with polarization directed along $[110]$ directions with respect to the cubic perovskite substrate \cite{Eklund09}. PbTiO$_{3}$ itself can develop in-plane polarization when grown on DyScO$_{3}$ \cite{Catalan06}, which exerts a tensile strain of around 1.4$\%$. Additionally, recent first principles predictions for the related superlattice system PbTiO$_{3}$/SrTiO$_{3}$ suggest it could have a continuous rotation of polarization from [001] to [110] as the strain is varied from compressive to tensile \cite{AguadoPuente11}. In our study, rather than modify strain, which is experimentally limited to certain discrete values by the choice of available substrates, we have prepared superlattice samples in which the constituent layers have different relative thicknesses. The aim of this study was to determine whether, for particular ratios of the constituent layer thicknesses, the competing tendencies of PbTiO$_{3}$ and CaTiO$_{3}$ would result in the overall direction of polarization being directed along one of the intermediate monoclinic directions associated with polarization rotation. In this study the CaTiO$_{3}$ layer thickness was fixed at 3 unit cells and all of the superlattices were epitaxially constrained in-plane to the SrTiO$_{3}$ substrate, both conditions which should encourage uniform polarization from layer to layer. The degree to which adjoining layers are coupled is open to question \cite{Johnston05,AguadoPuente12,TorresPardo11,Zubko12}, but not considered in detail here. 
 
PbTiO$_{3}$/CaTiO$_{3}$ superlattices were deposited using off-axis RF magnetron sputtering (conditions in Supplemental Material) on (001) SrTiO$_3$ substrates, which had been treated with buffered HF and annealed to ensure TiO$_{2}$ termination. In each of the superlattices discussed here the CaTiO$_{3}$ layer thickness was 3 unit cells and the PbTiO$_{3}$ thickness was varied from sample to sample. Here we consider changes in properties as a function of the CaTiO$_{3}$ volume fraction, $x=\frac{n_{CaTiO_{3}}}{n_{CaTiO_{3}}+n_{PbTiO_{3}}}$ where $n_{CaTiO_{3}}$ and $n_{PbTiO_{3}}$ are the number of unit cells of each material in the superlattice bilayer.   Most of the samples discussed here also had bottom SrRuO$_{3}$ electrodes (20 nm in thickness). Gold top electrodes were added to the samples post-deposition. For $x\leq0.5$ the total thickness of each superlattice was 100nm.  For superlattices with $x>0.5$ the epitaxial constraint from the substrate could only be maintained by reducing the total thickness of the superlattice, but as this led to high leakage currents we do not show electrical measurements for these samples. High Resolution-Scanning Transmission Electron Microscopy (HR-STEM) and x-ray diffraction both reveal that the samples are grown epitaxially constrained to the in-plane lattice parameter of the substrate, that the layers in the structure are well defined with no evidence of grain boundaries or misfit dislocations (see Supplemental Material).

\begin{figure}
  \includegraphics[width=8.6cm]{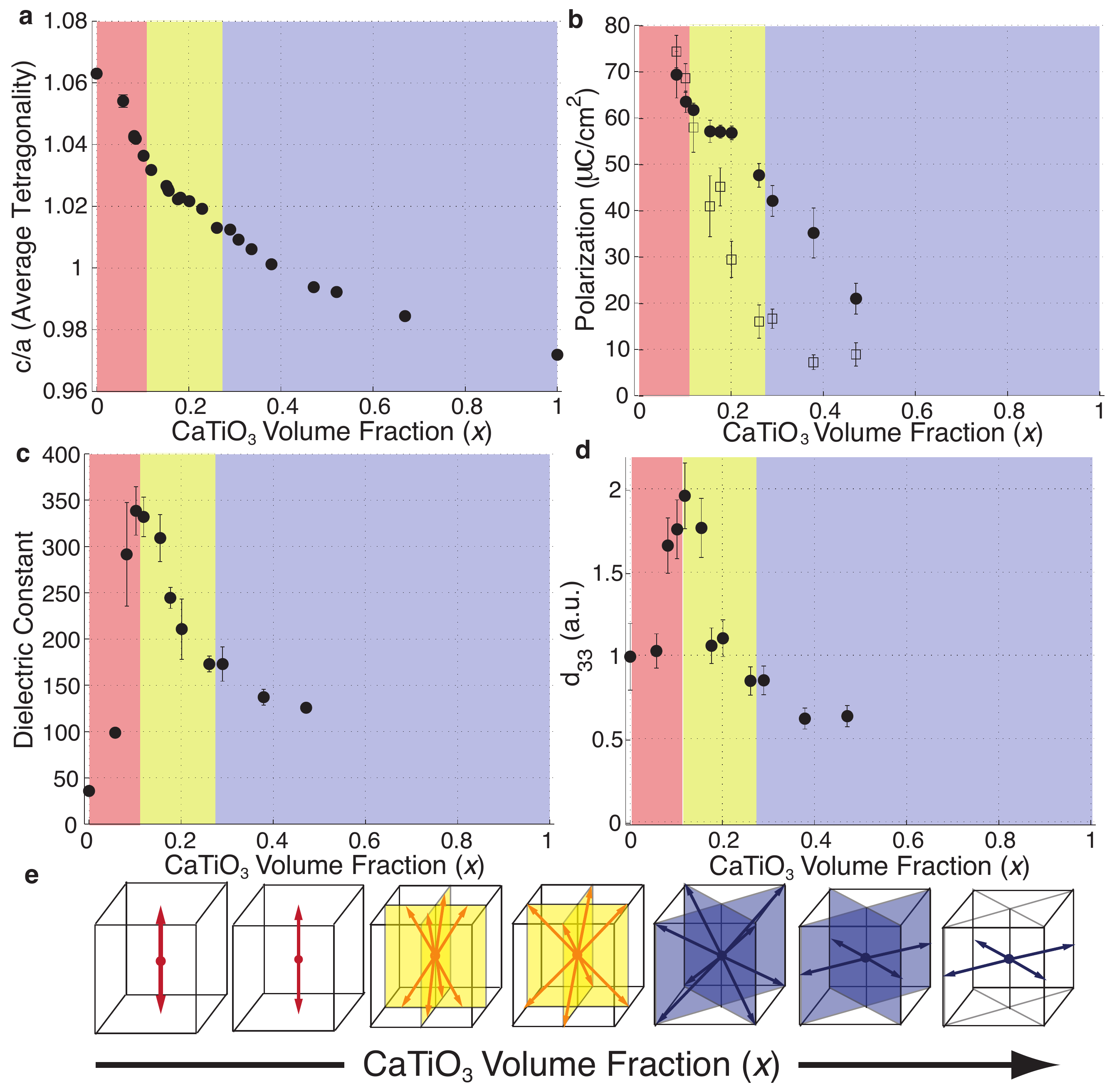}
  \caption{\textit{(a-d), c/a (tetragonality), polarization, dielectric constant and d$_{33}$ plotted as a function of CaTiO$_{3}$ volume fraction ($x$).  The general evolution of the polarization in these regions, from [001] to [$u0v$] to [$uuv$] is depicted schematically in (e).}}\label{4plots}
\end{figure}

\begin{figure}
  \includegraphics[width=8.6cm]{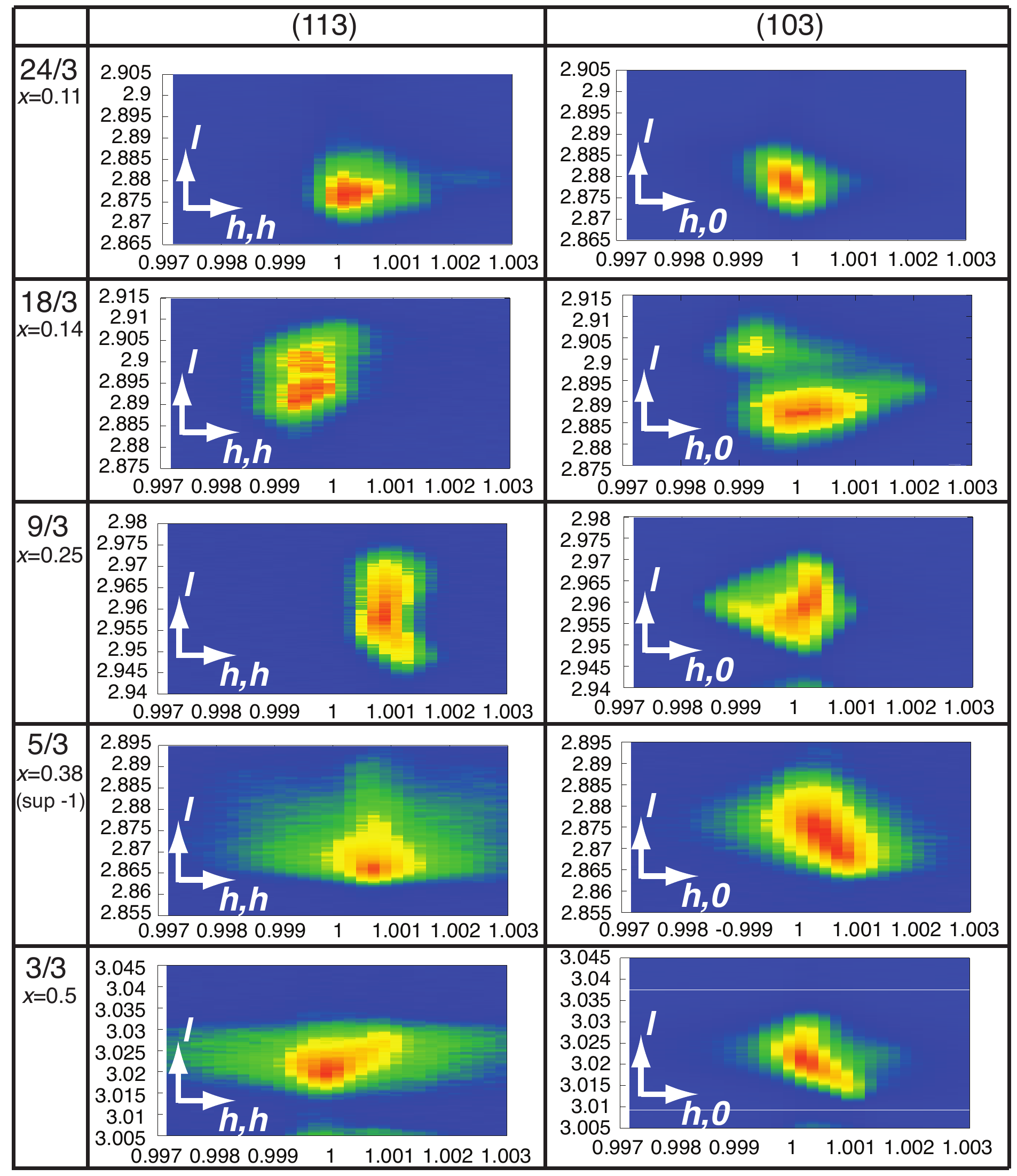}
  \caption{\textit{X-ray diffraction reciprocal space maps around the $(113)$ and $(103)$ peaks of  5 PbTiO$_{3}$/CaTiO$_{3}$  superlattice samples. The x and y axes are labelled with respect to the SrTiO$_{3}$ lattice. Intensity is plotted on a linear color scale with intensity represented from low to high by: blue-green-yellow-red.). Note that as the main superlattice peaks of the 5/3 sample lie directly on top of the much more intense SrTiO$_{3}$  substrate peak the scan shown is on the -1 superlattice satellite peak. It was verified on other samples that superlattice satellite peaks show the same kind of splitting as the main superlattice peaks. The measurements shown were performed on a Bruker D8 Discover diffractometer with Cu K-$\alpha$ radiation.} }\label{113103maps}
\end{figure}

The average tetragonality of each superlattice sample was measured using x-ray diffraction, using a Bruker D8 Discover diffractometer (Cu K-$\alpha$ radiation). The tetragonality was obtained  from the (001) and (002) peaks of the superlattice using simulations which took in to account the underlying SrRuO$_{3}$ electrode. Values were confirmed by measurements of the position of the (113) peak,  a peak for which SrRuO$_{3}$ has little or no intensity. Samples of the same composition, both with and without bottom electrodes, were found to have the same tetragonality. The tetragonality of the samples is plotted as function of the CaTiO$_{3}$ volume fraction, $x$, in Fig. \ref{4plots} (a). There is noticeable similarity between the curve obtained here and that for pure PbTiO$_{3}$ as a function of pressure, shown in Fig. 3 of Ahart et al. \cite{Ahart08}. Confirmation that the changes in slope of the tetragonality vs CaTiO$_{3}$ volume fraction are associated with changes of symmetry is provided by x-ray diffraction reciprocal space maps around the $(113)$ and $(103)$ peaks, which show different kinds of splitting for $M_{C}$, $M_{A}$, $R$ and $O$ phases \cite{Xu05, Christen11}.  The results on our samples (Fig. \ref{113103maps}) show splitting compatible with a monoclinic distortion along $[u0v]$ for samples with $0.13<x<0.26$. For samples with higher CaTiO$_{3}$ volume fraction there appears to be a smooth transition through a $M_{A}$ phase with polarization along $[uuv]$  towards an orthorhombic structure distorted along [110].

\begin{figure}[h]
  \includegraphics[width=8.6cm]{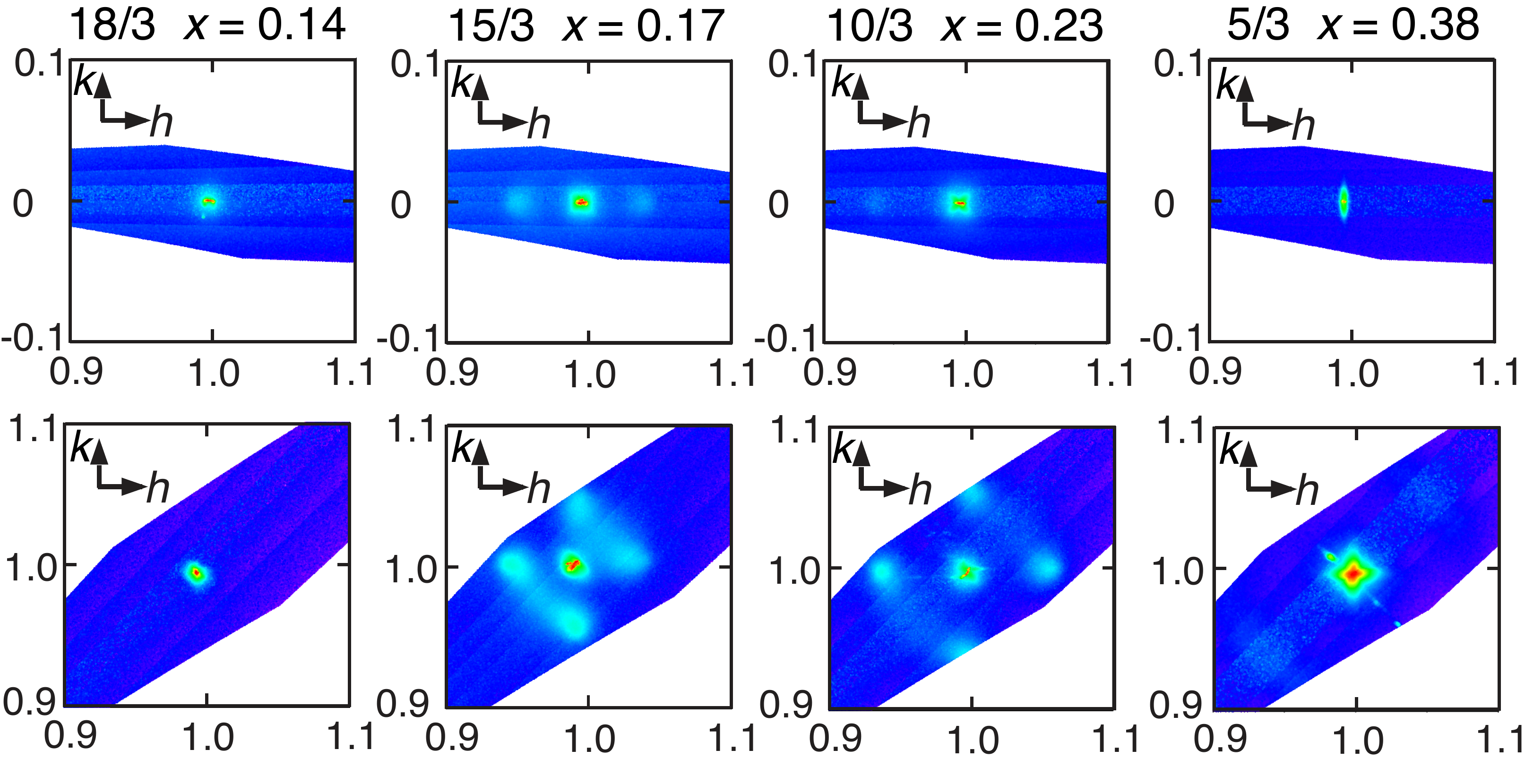}
  \caption{\textit{Grazing incidence x-ray diffraction reciprocal space maps along in-plane directions around the substrate (100) (top) and (110) (bottom) peaks for from left to right, 18/3, 15/3, 10/3 and 5/3 PbTiO$_{3}$/CaTiO$_{3}$ superlattices. Intensity is plotted on a logarithmic color scale with intensity represented from low to high by: blue-green-yellow-red.)}}\label{xraymaps}
\end{figure}

In addition to the scans shown in Fig.\ref{113103maps}, lower resolution scans were made over a broader range of reciprocal space (shown in Supplemental Material). All the samples with volume fractions $x<0.26$ show satellite peaks around $(113)$ indicative of domains, which appear to gradually decrease in size as the CaTiO$_{3}$ volume fraction is increased. Satellite peaks around $(113)$ can occur for domains which are either entirely in-plane, entirely out-of-plane, or somewhere in between. To confirm that an in-plane polarization developed in the films we performed in-plane grazing incidence x-ray diffraction around $(hk0)$ substrate peaks \cite{Fong04,Catalan06}. If scattering from domains is seen around these peaks it shows that there is an in-plane component of the polarization.  Grazing incidence in-plane x-ray diffraction was performed at  beamlines X22C and X21 at the National Synchrotron Light Source at Brookhaven National Laboratory. The x-ray radiation used in the synchrotron measurements had energy 10keV. In addition to the scans around (100) and (110)  shown, scans were also made around (010) and (210), and the full set of scans was made on a total of 8 samples. The results shown here were obtained at X21 using a Pilatus area detector to rapidly map out large areas of reciprocal space with high resolution. Comparable results were obtained using a point detector at X22C, but, as those maps have a lower point density due to additional time required to perform the scans, we have presented the maps obtained at X21. In Fig. \ref{xraymaps} we show reciprocal space maps made around the $(100)$ (top row) and $(110)$ (bottom row) substrate peaks for four samples.  No in-plane features are seen for samples with $x<0.15$.  For samples with  $0.15<x <0.26$ spots  appear along [110] directions,  and these are replaced by weaker spots along $[010]$ directions for $x>0.26$. These features are compatible with scattering from domain structures associated with polarizations that have some component of their polarization directed along the direction in which the features are seen. Similar features were observed under applied electric field in PZN-PT by Noheda et al \cite{Noheda01}. Here they are observed in the absence of an applied field because the c axis direction is fixed by the epitaxial strain coming from the SrTiO$_{3}$ substrate.  As the CaTiO$_{3}$ volume fraction is increased, the domain features are seen increasingly far from the central diffraction spot (in agreement with the domain features observed around (113) peaks); this implies that the in-plane domain spacing becomes smaller as the polarization starts to rotate into the plane. A close inspection of the scans around $(100)$ for the 15/3 and 10/3 samples show that, in addition to the well defined spots far from the central substrate spot, there is ``butterfly-like'' diffuse scattering in the immediate vicinity of the substrate peaks, which is similar to that observed in PMN-PT  \cite{Xu04PMN} and  PZN-PT \cite{Xu04PZN}. For the 5/3 sample, in which polarization has rotated to be in the $[uuv]$ plane, these features have also rotated, ie. the ``butterfly'' is now seen around the $(110)$ peak (and is rotated by 45$^{o}$) and the (100) scan now shows ellipsoid diffuse scattering, similar to that normally seen around the $(110)$ peak in rhombohedral PMN-PT \cite{Xu04PMN} and  PZN-PT \cite{Xu04PZN},  though this scattering is  also rotated by 45$^{o}$ when compared to those materials.  

A representation of the general changes in polarization directions as one progressively increases the CaTiO$_{3}$ volume fraction is shown in Fig. \ref{4plots} (e). The observation of a phase with polarization along $[u0v]$ in our samples is particularly interesting. While the transition through two monoclinic phases is seen in pure PbTiO$_{3}$ under pressure at low temperatures, the solid solutions generally only show one monoclinic phase, and the first principles predictions for polarization rotation in PbTiO$_{3}$/SrTiO$_{3}$ superlattices under increasing tensile strain \cite{AguadoPuente11} pointed towards rotation from [001] to [110], with no intermediate $[u0v]$ phase.  There have been a number of studies regarding the  phase diagram of the Pb$_{1-x}$Ca$_{x}$TiO$_{3}$ solid solution, but there is not universal agreement on the phases that exist between the tetragonal structure seen from $0\leq x<0.4$ and orthorhombic CaTiO$_{3}$, with some studies finding a cubic intermediate phase\cite{Kuo04}, and others pointing towards various lower symmetry phases, none of which has been identified as monoclinic, between the two clearly defined endpoints\cite{Sawaguchi59,Ganesh97,Torgashev06}. In any case, our samples differ substantially from the solid-solution, both because of their highly ordered structure and the progressive change from highly compressive to highly tensile strain as the CaTiO$_{3}$ volume fraction is increased. We also note that in the composition range where we identify polarization rotation through a monoclinic phase the solid solution has been clearly identified as tetragonal.

Having demonstrated that the structural changes which allow polarization rotation occur in our materials, characterization of their functional properties was performed.   In Fig \ref{4plots}. (b) values for of both the switched polarization (solid circles) and remnant polarization (open squares) are plotted as a function of CaTiO$_{3}$ volume fraction. (Details on the method used to obtain these values, and characteristic P-E loops, can  be found in the Supplemental Material).  It can be seen that within the $M_{C}$ phase in our materials, there is little change in the the magnitude of the switched polarization with volume fraction, in contrast to the steady decrease observed at other compositions. As the CaTiO$_{3}$ volume fraction increases, the polarization become progressively less stable, leading to the observed decrease in the remnant polarization.

\begin{figure}[h]
  \includegraphics[width=8.6cm]{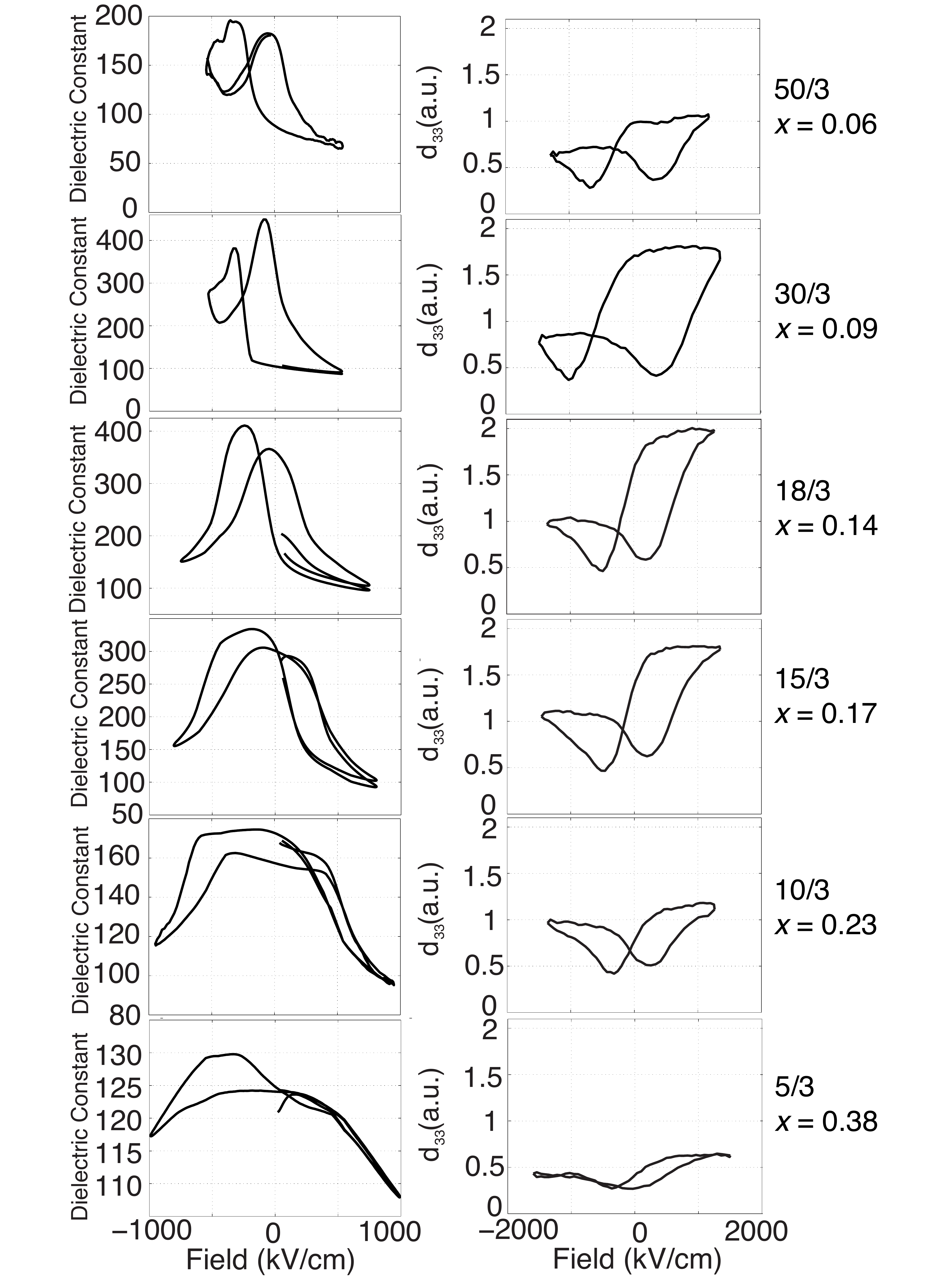}
  \caption{\textit{Dielectric and piezoelectric susceptibility measurements performed on, from top to bottom, 50/3, 30/3, 18/3, 15/3, 10/3 and 5/3 PbTiO$_{3}$/CaTiO$_{3}$ superlattices. Left, dielectric constant-electric field loops, and right, d$_{33}$ (in arbitrary units) as a function of applied bias field measured using piezoforce microscopy.} }\label{functionalmeasurements}
\end{figure}

The dielectric constant of the samples was measured as a function of DC bias using an LCR meter, (for details, see Supplemental Material).  The dielectric constant value shown in Fig. \ref{4plots} (c) was obtained for each sample from the crossing point of the traces in the butterfly loops that were measured (examples are shown in Fig. \ref{functionalmeasurements}). The dielectric constant peaks at the transition to $[uuv]$ oriented polarization and has a small plateau, before rapidly falling off in the $M_{C}$ phase, and becoming fairly constant for $x>0.26$. The relatively constant polarization in the $M_{C}$ phase combined with the enhanced dielectric constant in the vicinity of the $T-M_{C}$ transition is a good indication that piezoresponse should be high in these samples, and the highly tunable dielectric properties in this region are attractive by themselves.

Finally, the out-of-plane strain generated in response to an out-of-plane field ($d_{33}$) was measured directly using DFRT-PFM \cite{Rodriguez07} on an Asylum MFP-3D atomic force microscope  (Fig. \ref{functionalmeasurements}). The measurement was performed using a Au coated AFM tip (MikroMasch NSC18/Cr-Au/50) as the top electrode, which was applied directly to the superlattice top layer, not the gold top electrode used for electrical characterization. Due to numerous experimental factors the absolute magnitude of this piezoresponse is challenging to determine, so the values plotted are the maximum values of the piezoresponse for each sample normalized by the highest measured piezoresponse for a pure PbTiO$_{3}$ film (which by other means has been measured to be $\approx 60$pm/V\cite{Stucki}). These values correspond to the right hand side of the measured loop, which consistently has higher values than the left, even for the pure PbTiO$_{3}$ film. The asymmetry is most likely due to the difference between the top and bottom electrodes (gold coated AFM tip on top and SrRuO$_{3}$ film on the bottom).  The enhancement of the piezoresponse as a function of CaTiO$_{3}$ volume fraction is seen whether the values are taken from the right or the left side of the loop.  Although domain wall motion can contribute to high piezoresponses it is believed that the enhanced piezoresponse in the these samples comes from polarization rotation rather than domain wall motion, as here the spacing of the domains changes smoothly in the vicinity of the sharp peak in the piezoelectric and dielectric responses (see reciprocal space maps in Supplemental Material). 

Our results suggest a path forward to the development of new piezoelectric materials. The enhancement of piezoresponse here is driven by the competition between preferred directions for polarization between two materials in a layered structure.  As these samples are thin films, high electric fields can be readily applied, and the ordered layered structure that these samples have is a far simpler system to study by density functional theory methods than solid solutions which require very large supercells.  We thus expect that this material system will act as an excellent model system to help understand polarization rotation behavior in both pure single crystals and solid solutions. 

This work was supported by the National Science Foundation under DMR1055413 ``CAREER: Engineered Ferroic Superlattices for Science, Technology and Education''. Use of the National Synchrotron Light Source and the Center for Functional Nanomaterials, at Brookhaven National Laboratory, was supported by U.S. Department of Energy, Office of Basic Energy Sciences, under Contract No. DE-AC02-98CH10886.  MD and DS  acknowledge support from a SBU/BNL seed grant.

\end{document}